\documentclass[%
 aps,
 prl,
 amsmath,amssymb,
 reprint,
]{revtex4-1}

\usepackage{graphicx}
\usepackage{dcolumn}
\usepackage{bm}

\usepackage{color}

\usepackage{float}  

\begin{document}


\title{Time-dependent density functional theory for a unified description of ultrafast dynamics:
  pulsed light, electrons, and atoms in crystalline solids}

\author{Atsushi Yamada}
\email{ayamada@ccs.tsukuba.ac.jp}
\author{Kazuhiro Yabana}%
\affiliation{%
Center for Computational Sciences, University of Tsukuba, 1-1-1 Tennodai, Tsukuba, Ibaraki 305-8577, Japan 
}%

%
%

\date{\today}

\begin{abstract}
We have developed a novel multiscale computational scheme to describe coupled dynamics of light electromagnetic field 
with electrons and atoms in crystalline solids, 
where first-principles molecular dynamics based on
time-dependent density functional theory is used to describe the microscopic dynamics. 
The method is applicable to wide phenomena in nonlinear and ultrafast optics.
To show usefulness of the method, we apply it to a pump-probe measurement of coherent phonon in diamond
where a Raman amplification takes place during the propagation of the probe pulse.
\end{abstract}

\pacs{Valid PACS appear here}
\maketitle



Nonlinear optics in solids is the study of the interaction of intense laser light with
bulk materials \cite{Boyd2008book,Shen1984book,Bloembergen1996book}.
It is intrinsically a complex phenomena arising from a coupled nonlinear 
dynamics of light electromagnetic fields, electrons, and phonons. 
They are characterized by two different spatial scales, micro-meter for the 
wavelength of the light and less than nano-meter for the dynamics of electrons and atoms.

In early development, nonlinear optics has developed mainly in perturbative regime 
and in frequency domain\cite{Goodman2005book,Hecht2002book}. However, it has changed rapidly and drastically. 
Nowadays, measurements are carried out quite often in time domain using pump-probe 
technique as a typical method and the time resolution reaches a few tens of attosecond\cite{Krausz2009,calegari2016}. 
Extremely nonlinear phenomena has attracted interests such as high harmonic 
generation in solids\cite{Ghimire2011,Luu2015}, ultrafast control of electrons motion in dielectrics that aims for
future signal processing using pulsed light\cite{Schultze2014,Sommer2016,Lucchini2016},
and ultrafast coherent optical phonon control\cite{Hase2003,Mizoguchi2013,Kitajima2013,Takeda2014,Nakamura2016,Nakamura2018,Merlin1997,Merlin2002,Thomsen1984,Cho1990}
and photoinduced structural phase transition
of materials\cite{Okamoto2017,Horiuchi2017,Schmidt2017,Marieke2017}.

In this paper, we report a progress to develop first-principles computational method
to describe nonlinear optical processes in solids that arise from coupled dynamics of 
light electromagnetic fields, electrons, and atoms in crystalline solids.
In condensed matter physics and materials sciences, first-principles computational 
approaches represented by density functional theory have been widely used and 
recognized as an indispensable tool\cite{Martin2004book}. Development of first-principles 
approaches in optical sciences is, however, still in premature stage due to the complexity 
of the phenomena and the requirement of describing time-dependent dynamics.

Our method utilize time-dependent density functional theory (TDDFT) for microscopic
dynamics of electrons\cite{Ullrich2012, Marques2012}.
The TDDFT is an extension of the density functional theory
so as to be applicable to electron dynamics in real time\cite{Runge1984}. 
In microscopic scale, ultrafast dynamics of electrons
have been successfully
explored solving the time-dependent Kohn-Sham (TDKS) equation, the basic equation of TDDFT,
in real time\cite{Yabana1996, Otobe2008, Shinohara2010}.
We have further developed a multiscale scheme coupling the light electromagnetic fields and 
electrons in solids\cite{Yabana2012}, and applied it to investigate extremely nonlinear optical 
processes in dielectrics using few-cycle femtosecond pulses\cite{Sommer2016, Lucchini2016}. 
Here we extend the approach to incorporate atomic motion,
combining first-principles molecular dynamics approach\cite{Curchod2013}. 
It will then be capable of describing 
vast nonlinear optical phenomena involving atomic motion such as stimulated Raman 
scattering\cite{Boyd2008book,Shen1984book,Bloembergen1996book} that will be discussed later in this paper.

\begin{figure}[t]
\centering
\includegraphics[angle=0,width=0.98\linewidth]{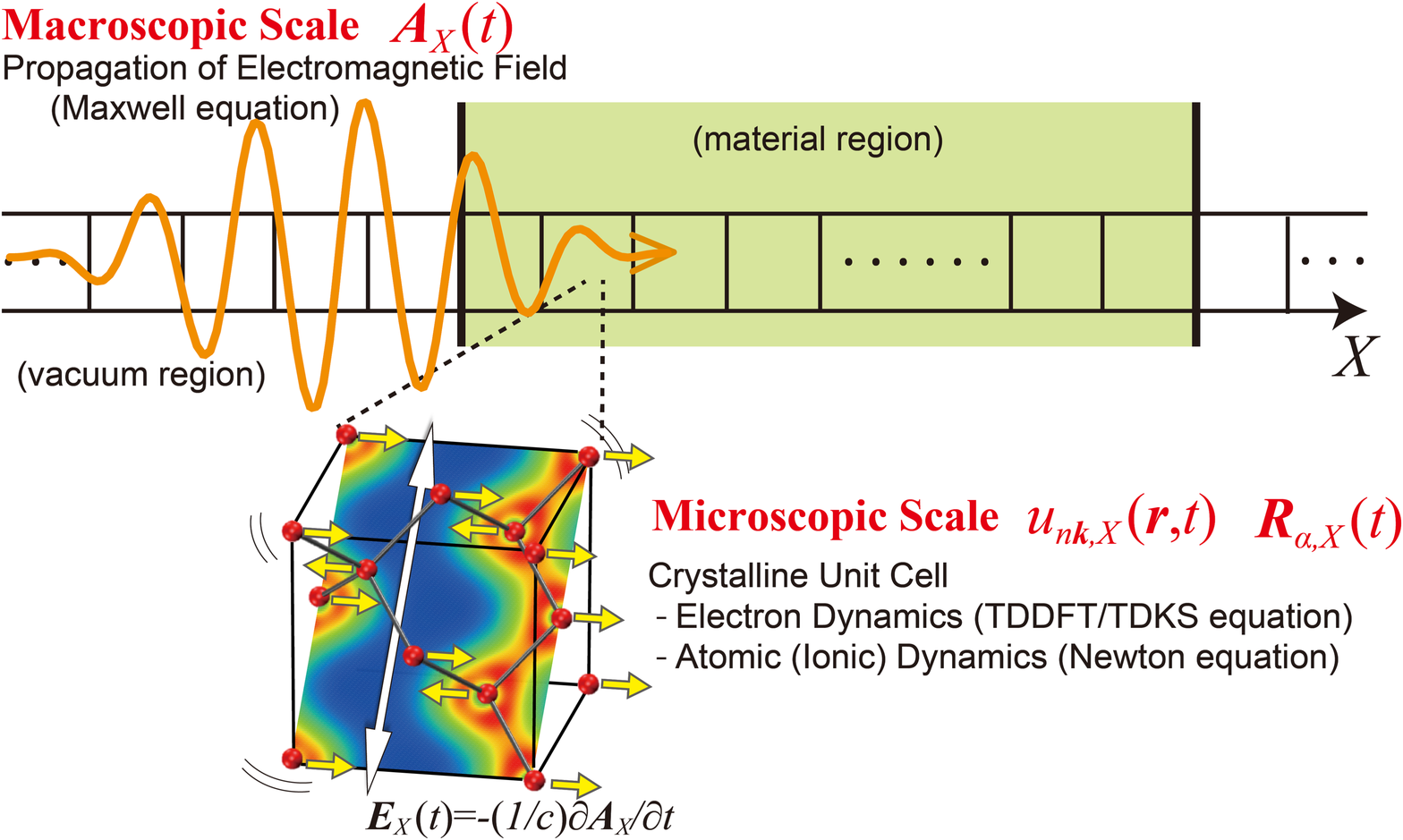}
\vspace{-4mm}
\caption{
  Schematic illustration of the multiscale scheme for a light propagation through a material. 
}
\label{fig-simulation-method}
\end{figure}

We consider a problem of an ultrashort pulsed light irradiation normally on a thin
dielectric film, as illustrated in Fig.\ref{fig-simulation-method}.
In our multiscale description \cite{Yabana2012}, we introduce two coordinate systems, 
the macroscopic coordinate $X$ that is used to describe the propagation of the light 
electromagnetic fields, and the microscopic coordinates $\bm r$ for the dynamics of 
electrons and atoms.

The light electromagnetic field is expressed by using a vector potential
${\bm A}_X(t)$ that satisfies the Maxwell equation,
\begin{equation}
\left[ \frac{1}{c^2} \frac{\partial^2}{\partial t^2} - \frac{\partial^2}{\partial X^2} \right] {\bm A}_X(t)
= \frac{4\pi}{c} {\bm J}_X(t),  \label{Maxwell}
\end{equation}
where ${\bm J}_X(t)$ is the electric current density at the point $X$.

We will use a uniform grid for the coordinate $X$ to solve Eq.(\ref{Maxwell}).
At each grid point $X$, we consider an infinitely periodic medium composed of electrons and atoms
whose motion are described using the microscopic coordinate $\bm r$.
Since the wavelength of the pulsed light is much longer than the typical spatial scale of
the microscopic dynamics of electrons and atoms, we assume a dipole approximation:
At each point $X$, the electrons and atoms evolve under a spatially-uniform electric field,
$E_X(t) = -(1/c)(\partial A_X(t)/\partial t)$.
Then we may apply the Bloch theorem at each time $t$ in the unit cell,
and may describe the electron motion using Bloch orbitals $u_{n {\bm k},X}({\bm r},t)$ with the
band index $n$ and the crystalline momentum ${\bm k}$ \cite{Bertsch2000}.
Atomic motion can also be described by atomic coordinates in the unit cell,
${\bm R}_{\alpha,X}(t)$, where the index $\alpha$ distinguishes different atoms in the unit cell.
 
The Bloch orbitals satisfy the TDKS equation,
\begin{eqnarray}
&& i\hbar \frac{\partial}{\partial t} u_{n{\bm k},X}({\bm r},t)= \nonumber\\
&& \left[ \frac{1}{2m} \left\{ -i\hbar \bm \nabla_{\bm r} + \hbar{\bm k} + \frac{e}{c} {\bm A}_X(t) \right\}^2
-e \phi_X({\bm r},t) \right.  \nonumber\\
&& \left. + \frac{\delta E_{XC}[n_{e,X}]}{\delta n_{e,X}}
+ \hat v_{ion,X}(\bm r;\{\bm R_{\alpha,X}(t)\}) \right] u_{n{\bm k},X}({\bm r},t) \label{TDKS} 
\end{eqnarray}
where 
$n_{e,X}$ is the electron density given by $n_{e,X}({\bm r},t)=\sum_{n,{\bm k}}|u_{n{\bm k},X}({\bm r},t)|^2$.
$\phi_X({\bm r},t)$ and $E_{XC}[n_{e,X}]$ are the Hartree potential and the exchange-correlation energy, respectively. 
$\hat v_{ion,X}(\bm r; \{ \bm R_{\alpha,X}(t)\})$ is the electron-ion potential for which we use
norm-conserving pseudopotential \cite{Troullier1991}. 

To describe the dynamics of atoms, we use a so-called Ehrenfest dynamics \cite{Ullrich2012}
where the atomic motion is described with mean-field from the electrons
by the Newton equation,
\begin{eqnarray}
  M_{\alpha} \frac{d^2 \bm R_{\alpha,X}}{dt^2}
  =-\frac{eZ_{\alpha}}{c} \frac{d\bm A_X}{dt}
  +\frac{\partial}{\partial \bm R_{\alpha,X}}
\int d\bm r [en_{ion,X}\phi_X]  \label{Newton} 
\end{eqnarray}
where $M_{\alpha}$ is the mass of the $\alpha$-th ion, $n_{ion,X}$ is the charge density of ions given by
$n_{ion,X}({\bm r},\{\bm R_{\alpha,X}(t)\})=\sum_{\alpha}Z_{\alpha}\delta({\bm r}-{\bm R_{\alpha,X}}(t))$, 
with $Z_{\alpha}$ the charge number of the $\alpha$-th ion. 

The electric current density at point $X$, $\bm J_X(t)$, consists of electronic and ionic contributions,
\begin{equation}
\bm J_X(t) = \bm J_{e,X}(t) + \bm J_{ion,X}(t).
\label{current}
\end{equation}
The electronic component $\bm J_{e,X}(t)$ is given by the Bloch orbitals $u_{n\bm k,X}(\bm r,t)$ \cite{Yabana2012}.
The ionic component is given by
$\bm J_{ion,X}(t) = e\sum_{\alpha} Z_{\alpha} (d{\bm R}_{\alpha,X}(t)/dt)/\Omega$, where $\Omega$ is the volume of the unit cell.

We solve Eqs. (\ref{Maxwell}) - (\ref{current}) simultaneously to obtain the whole dynamics at once
with the initial condition that the electronic state at each point $X$ is set to the ground state solution of
the static density functional theory, atomic positions are set to their equilibrium
position in the electronic ground state, and the vector potential of the incident pulsed
light is set in the vacuum region in front of the thin film.
We note that the present scheme naturally includes the ordinary macroscopic electromagnetism in a weak
field limit, since solving the TDKS equation (\ref{TDKS}) is equivalent to utilizing the linear constitutive relation 
for a weak field.

Now, we apply the method to describe the pump-probe measurement that aims to investigate the
generation of coherent optical phonons \cite{Merlin1997}.
We consider a diamond thin film of 6 $\mu$m thickness, and two linearly-polarized pulses are irradiated 
successively and normally on the surface.
We set the crystalline $abc$ axes of cubic diamond to coincide with the $xyz$ axes of the Cartesian coordinates, respectively.

We implemented our scheme in the open-source software SALMON \cite{SALMON_paper2018, SALMON_web} which utilizes real-space uniform
grid representation to express orbitals.
Since a number of microscopic dynamics are calculated simultaneously in the multiscale scheme, we use a massively
parallel supercomputer with an efficient implementation of parallelization \cite{Hirokawa2016}. 
Adiabatic local density approximation is used for the exchange-correlation potential \cite{Perdew1981}.
The $X$ coordinate from $X$=0 to 6 $\mu$m is discretized by 400 grid points with the spacing of 15 nm. 
The microscopic unit cell of diamond consists of eight carbon atoms in the cubic cell of the side 
length of 3.567 {\AA}. The Bloch orbitals are expressed using $16^3$ uniform spatial grids in the unit cell
and 12$^3$ of $k$-points in the Brillouin zone. The time step is chosen as 2 as.

\begin{figure}[t]
\centering
\includegraphics[angle=0,width=0.87\linewidth]{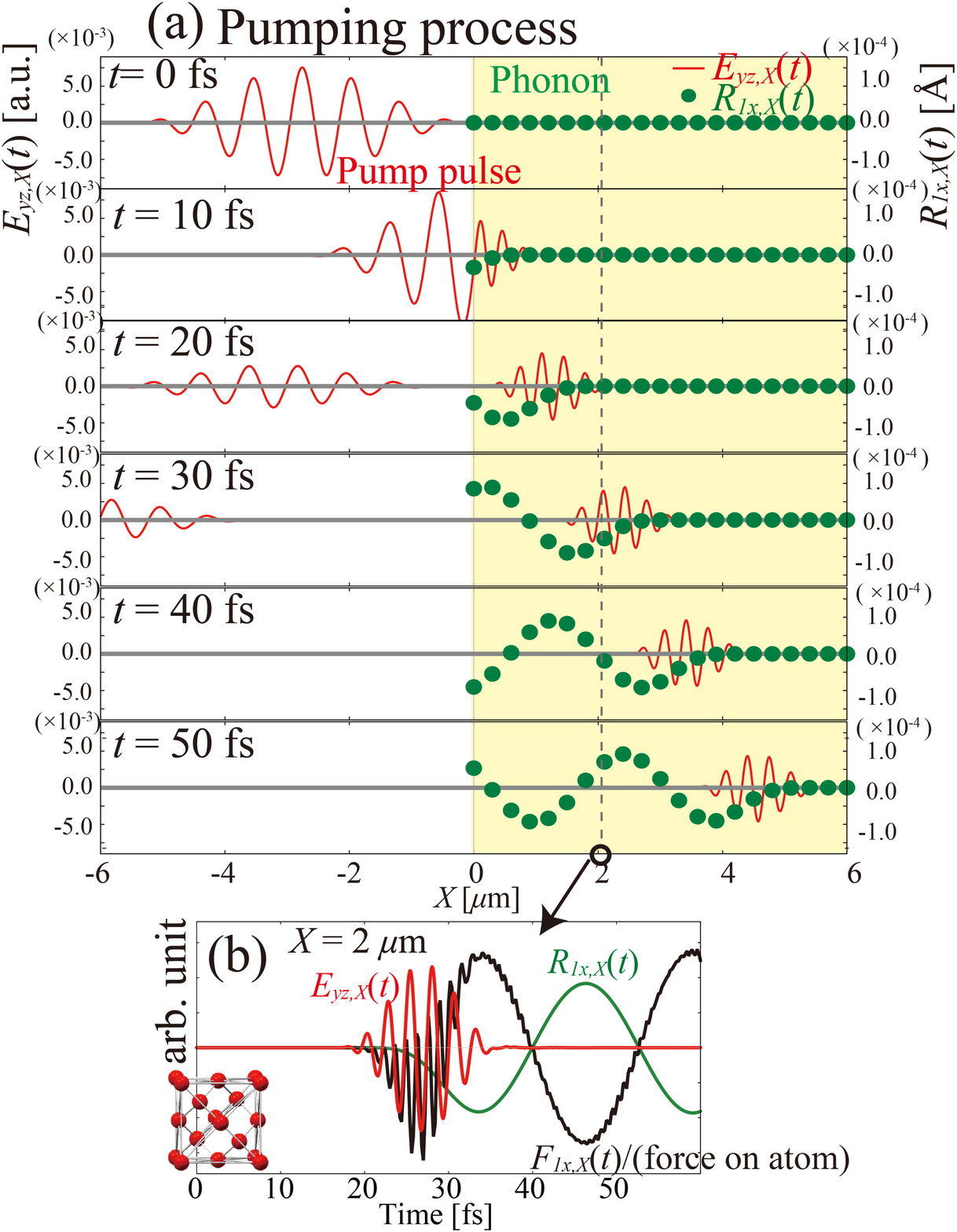}
\vspace{-4mm}
\caption{
  (a) Snapshots of the electric field of the pump pulse (red lines) and the atomic displacement (green filled circles) are shown 
  along the macroscopic position $X$.
  (b) Electric field of the pump pulse (red line), the force acting on the atom (black line), and the atomic displacement
  (green line) are shown at $X=2$ $\mu$m as a function of time.
  \label{fig-pump}}
\end{figure}

We first show the calculation for the generation of the coherent phonon by the pump pulse.
Figure \ref{fig-pump}(a) shows the propagation of the pump light and the displacement of 
the atoms at different macroscopic point $X$.
The frequencies and the pulse duration of the pump pulse is set to 1.55 eV$/\hbar$ and 6.5 fs in FWHM, respectively.
The pulse duration is chosen to be shorter than the period of the optical phonon of diamond, 25 fs.
The polarization direction of the pump pulse is chosen as [011] direction, which causes the optical phonon in
[100] direction. The maximum intensity of the pump pulse is set to $2\times 10^{12}$ W/cm$^2$.

When the pump field propagates through the material, the harmonic motion of atoms is generated in turn at each $X$ position. 
In Fig.~\ref{fig-pump}(a), a wave-like behavior of atomic displacement is seen along the $X$ axis.
However, this is not an ordinary phonon wave described by the lattice dynamics.
The period of the oscillation at each $X$ position is the period of the optical phonon which is 25 fs, 
and the speed of the propagation is equal to the speed of the pump pulse in the medium, $c/n$, with the index of refraction $n$.

In Fig.\ref{fig-pump}(b), the generation process of the optical phonon at $X$= 2 $\mu$m is shown as a function of time. 
At first, the force is proportional to the square of the electric field.
As the phonon starts to move, the restoring force begins to work. These behavior of the force and the generation process of the
coherent phonon is consistent with the picture of the impulsively stimulated Raman scattering (ISRS) mechanism\cite{Merlin1997}. 
After the pump field passes away, there is no driving force and a simple harmonic motion of atoms continues without decay.

\begin{figure}[t]
\centering
\includegraphics[angle=0,width=1.0\linewidth]{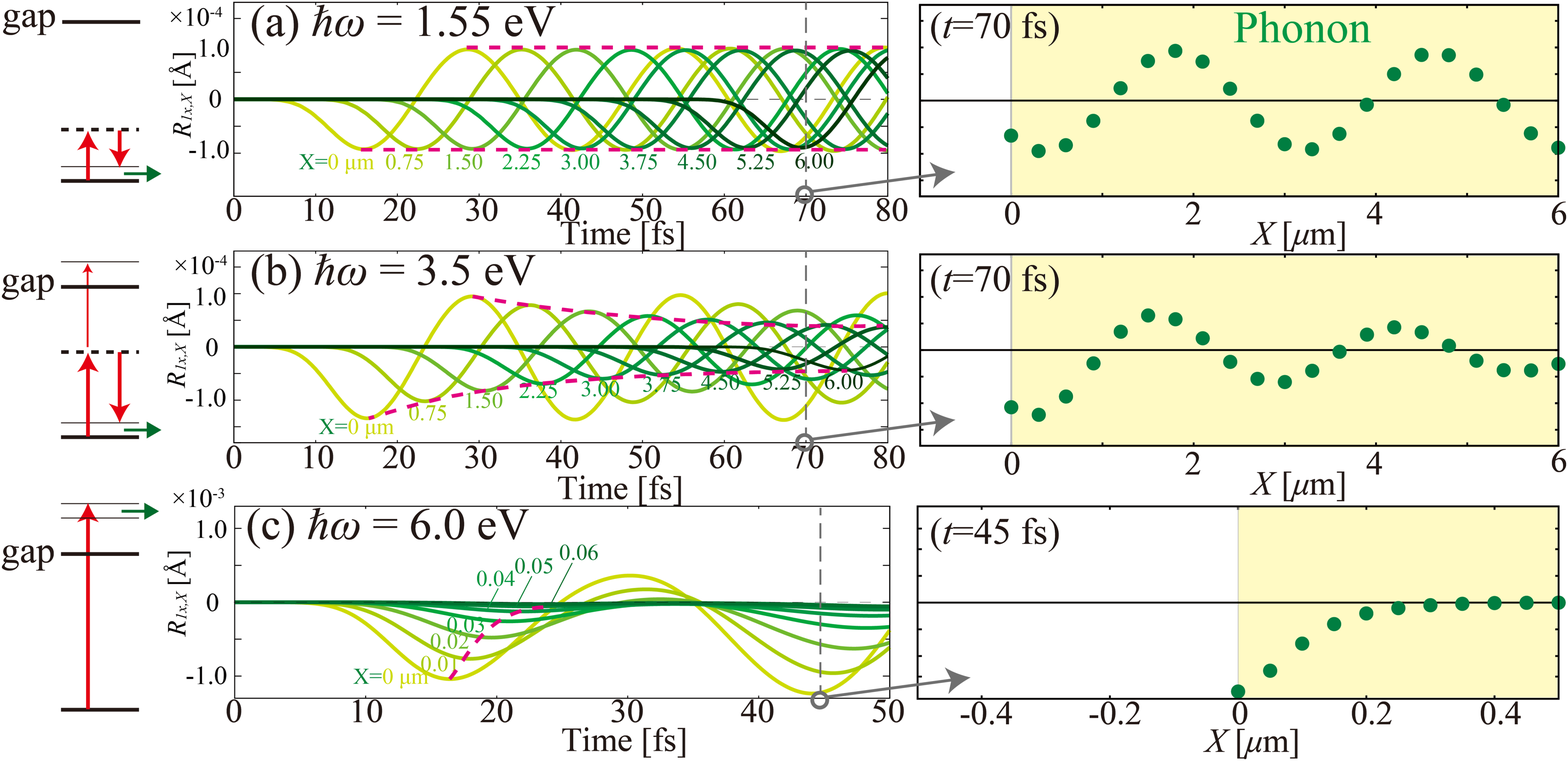}
\vspace{-4mm}
\caption{
  Coherent phonon generation by pump pulses of three different frequencies, $\hbar\omega$= (a)1.55, (b)3.5 and (c)6.0 eV.
  The atomic displacements as a function of time are shown at selected macroscopic positions of $X$ (the left panels)
  and those along the coordinate $X$ at specific times (the right panels).
  \label{fig-pump-large-omg}}
\end{figure}

We next show in Fig.\ref{fig-pump-large-omg} the generation of coherent phonons by pulses of three different frequencies.
Panel (a) shows the generation with $\hbar\omega$=1.55 eV, the same as that shown in Fig.\ref{fig-pump}(a).
Here the ISRS mechanism is responsible for the coherent phonon generation as mentioned above, since the frequency 
is below the optical gap energy.
At $\hbar\omega$=3.5 eV, the pump frequency is still below the optical gap energy. 
However, Fig.\ref{fig-pump-large-omg}(b) shows that the phonon amplitude decays with $X$ whereas the amplitude 
at each position $X$ does not decay with time. This is caused by the two photon absorption of the pump pulse.
Since the pump pulse is rather strong with the intensity, 2$\times$10$^{12}$ W/cm$^2$, the pump pulse 
excites electrons and looses energy as it propagate through the medium.

At $\hbar\omega$=6 eV that is above the optical gap energy, the generated phonon amplitude at the solid surface 
is one order of magnitude larger than the non-resonant cases, as seen in Fig.\ref{fig-pump-large-omg}(c), 
The displacement of atoms shows a harmonic motion of cosine shape, namely the oscillation takes place
around the shifted equilibrium position.
This is due to the change of the generation mechanism from ISRS to the 
displacive excitation of coherent phonon (DECP) \cite{Merlin1997}. The amplitude of the phonon decays with $X$ as the field gets weaker 
by the absorption of the pulse. 
These results are consistent with Ref. \cite{Shinohara2010} where generation mechanisms of coherent phonons
are discussed without describing the light propagation.

\begin{figure}[t]
\centering
\includegraphics[angle=0,width=0.98\linewidth]{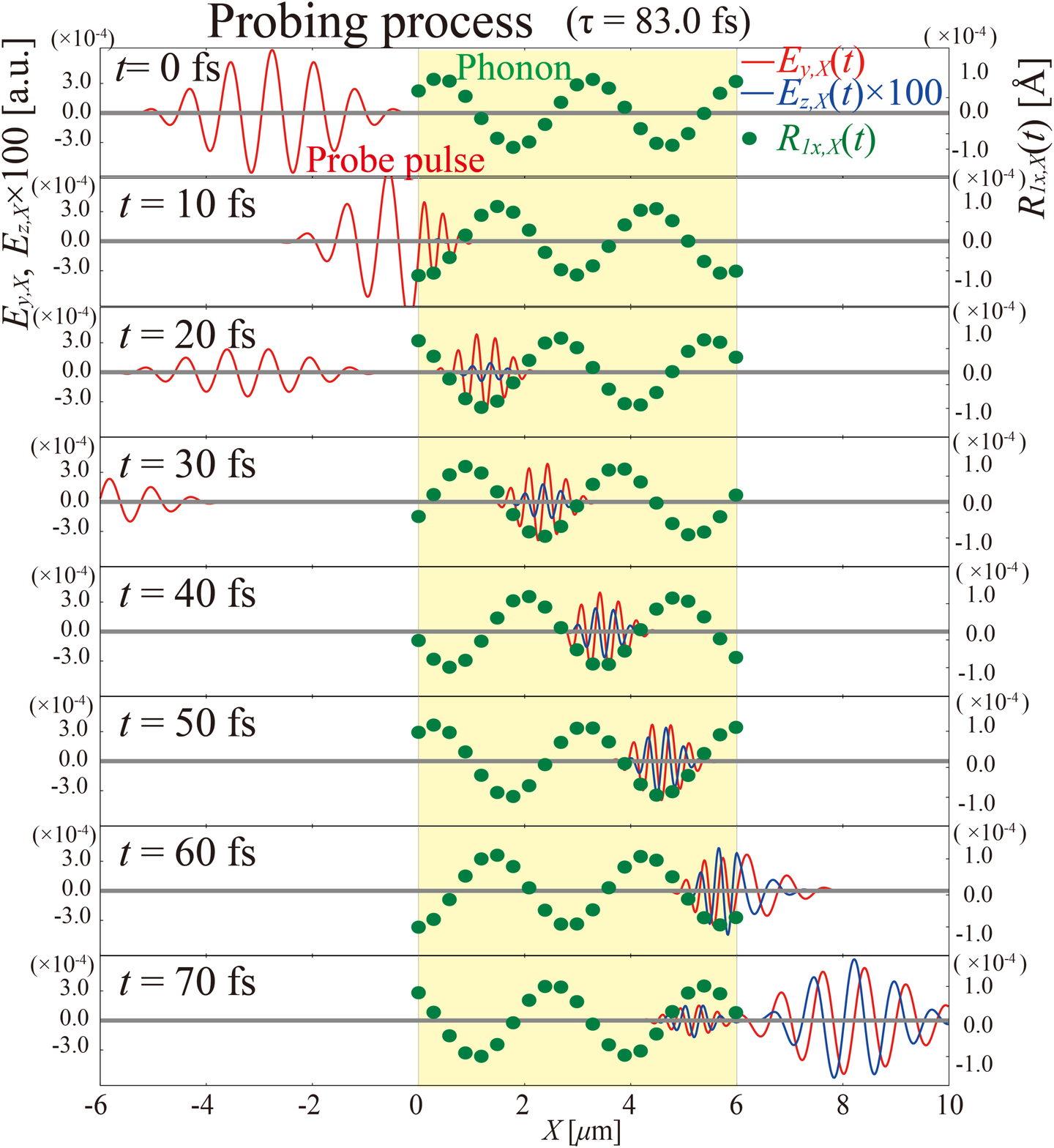}
\vspace{-4mm}
\caption{
  Snapshots of the electric field of the probe pulse in [010] (red-line)
  and in [001] (blue-line) polarization directions, and
 the atomic displacment (green filled circles) are shown along the macroscopic coordinate $X$.
  \label{fig-probe}}
\end{figure}

We next proceed to the calculation of the propagation of probe pulses.
Although it is possible to carry out the pump and the probe pulse propagations in a single calculation,
we separate them since the reflection of the pump pulse at the back surface of the thin film
complicates the analyses.
In the probe pulse calculation, we prepare the initial medium in which the coherent phonon already exists
and irradiate the probe pulse from the vacuum. The phonon motion is in [100] direction as discussed above, 
and we choose the polarization direction of the probe pulse as [010]. Then we expect the emergence and the
amplification of the stimulated Raman wave polarized along [001] direction due to the structure of the Raman tensor 
of the diamond \cite{Hu2008}.
The intensity of the probe pulse is set sufficiently small, $10^{10}$ W/cm$^2$, so that there occurs no significant nonlinearity
related to the probe pulse.

We show the propagation of the probe pulse and the amplification process of the stimulated Raman wave 
in Fig.\ref{fig-probe}.
In the figure, the probe pulse is irradiated with the pump-probe delay time of $\tau=83.0$ fs.
At this time delay, the center of the probe pulse moves with the maximum of the phonon amplitude,
as seen from the figure. Since, as we noted above, the phonon "wave" propagates with the speed of the
pump pulse in the medium and the speeds of the pump and the probe pulses are the same, the probe pulse 
always stay at the maximum of the phonon. 
During the propagation of the probe pulse in the medium, we find that the wave polarized along [001] direction
generated by the stimulated Raman scattering is amplified linearly in amplitude with the traveled distance
with the phase shift of $\pi/2$ relative to the probe pulse. These features are consistent with a description 
in standard theoretical description for the stimulated Raman scattering \cite{Merlin1997,Nelson1985}.

\begin{figure}[t]
\centering
\includegraphics[angle=0,width=0.85\linewidth]{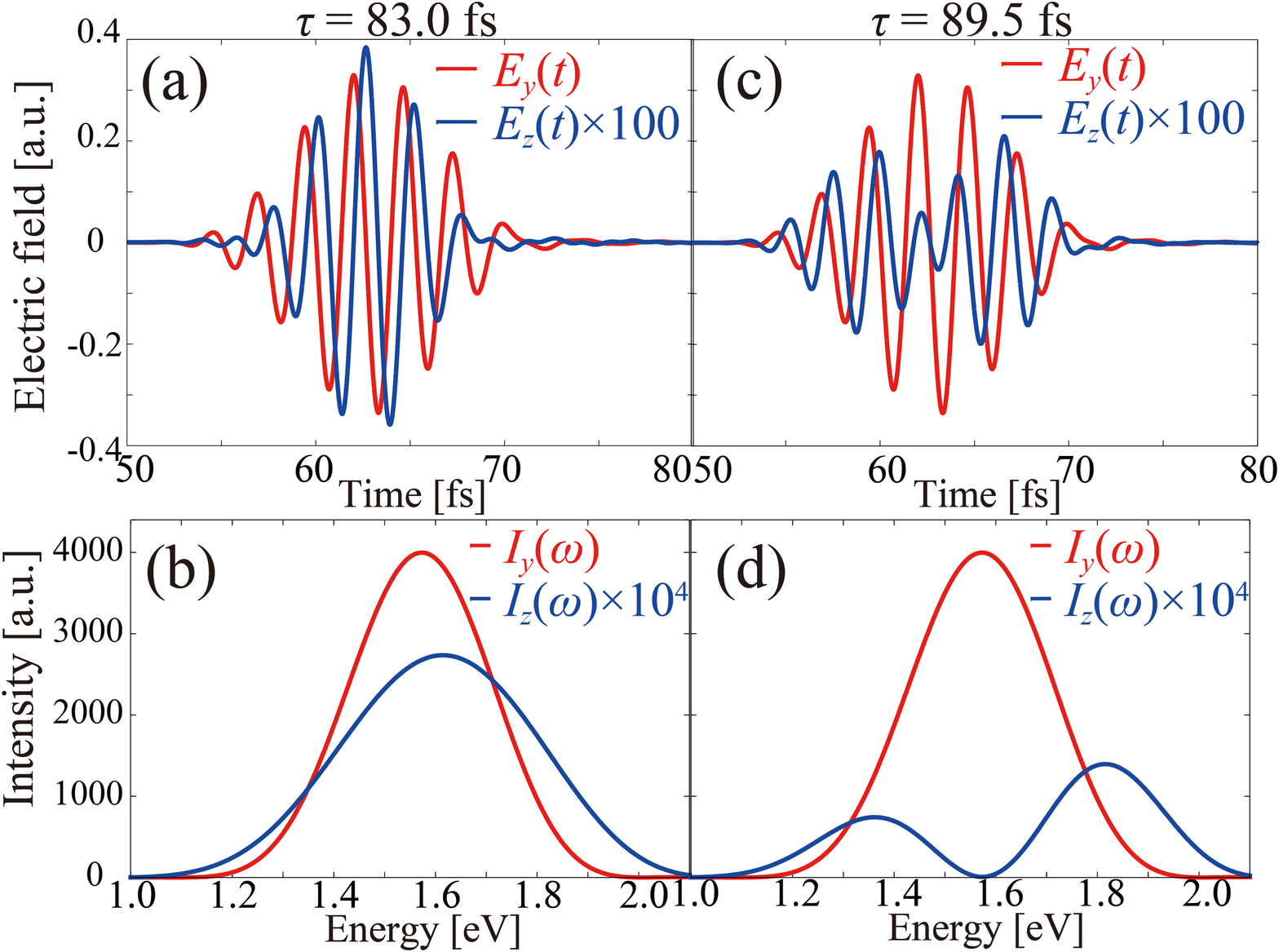}
\vspace{-4mm}
\caption{
  (a) Transmission wave in the right vacuum region and (b) the Fourier transformed power spectrum with $\tau$= 83.0 fs.
  Those with $\tau$= 89.5 fs are shown in (c) and (d).
  \label{fig-T_Et_Iw}}
\end{figure}
Finaly, we compare in Fig.\ref{fig-T_Et_Iw} the pulse shape of the probe pulse after it passes out
through the thin film for two different pump-probe time differences:
Panel (a) is the case of $\tau = 83.0$ fs, the same as that shown in Fig. \ref{fig-probe}.
Panel (b) is the case of $\tau = 89.5$ fs, at which the probe pulse moves with the nodal point 
of the phonon.

In the case of $\tau = 83.0$ fs, the pulse shape of the stimulated Raman wave is similar 
to the shape of the probe pulse, with the phase difference of $\pi/2$. On the other hand, 
in the case of $\tau=89.5$ fs, the pulse shape of the stimulated Raman wave is very different 
from the shape of the probe pulse.
This difference can be understood as being originated from the difference of the electric current 
density that produces the stimulated Raman wave. The electric current density that 
produce the Raman wave is given by $J_{\rm Raman}(t) \propto Q(t) E_{\rm probe}(t)$ \cite{Merlin1997,Nelson1985}, 
where $E_{\rm probe}(t)$ is the electric field of the probe pulse and $Q(t)$ is the phonon 
amplitude given by the linear combination of $\{ {\bm R}_{\alpha}(t) \}$.
When the probe pulse enters the medium at the maximum of the phonon amplitude, 
$Q(t)$ may be roughly regarded as a constant and $J_{\rm Raman}(t)$ is mostly proportional to $E_{\rm probe}(t)$
since the half period of the phonon is longer than the duration of the probe pulse. 
However, when the probe pulse moves with the nodal position of the phonon, the phonon amplitude 
is approximated by a linear function of time changing the sign.
Then, we have $J_{\rm Raman}(t) \propto t E_{\rm probe}(t)$,
and the phonon motion produces one extra node to the electric current density.
This explains the shape change of the stimulated Raman wave shown in Fig.\ref{fig-T_Et_Iw}(b).

We show the power spectra of the stimulated Raman waves in panels (b) and (d). 
Reflecting the difference in the time profile, the power spectra also show a distinct structures: 
the double-peak structure appears in the power spectrum of the stimulated Raman wave traveling
with the nodal point of phonon. We note that such double-peak structure is indeed related to
recent pump-probe measurement of the coherent phonon in diamond and other insulators \cite{Mizoguchi2013, Nakamura2016}.
We will report our analysis for this problem in a separate publication.

In summary, we have developed a computational approach for nonlinear light-matter
interaction in solids based on first-principles time-dependent density functional theory.
A multiscale scheme is developed simultaneously solving Maxwell equations for light propagation, 
time-dependent Kohn-Sham equation for electrons, and Newton equation for atoms.
As a test example, a pump-probe measurement of coherent phonon generation in diamond 
is simulated where an amplification by the stimulated Raman scattering is observed for the probe pulse.
We expect the method will be useful for a wide phenomena of nonlinear and ultrafast optics.

We acknowledge the supports by JST-CREST under grant number JP-MJCR16N5, and 
by MEXT as a priority issue theme 7 to be tackled by using Post-K Computer, 
and by JSPS KAKENHI Grant Number 15H03674. 
Calculations are carried out at Oakforest-PACS at JCAHPC through the Multidisciplinary 
Cooperative Research Program in CCS, University of Tsukuba, and through the HPCI System 
Research Project (Project ID: hp180088).

\bibliographystyle{unsrt}
\bibliography{bibf/yabana,bibf/theory,bibf/laser,bibf/tddft,bibf/coherent_phonon,bibf/other_misc_1}

\end{document}